\documentclass{aastex}
\usepackage{spr-astr-addons}
\usepackage{url}
\usepackage{graphicx}

\RequirePackage{color}

\begin{document}

\title{Electromagnetic interaction of a magnetized rotating star with
a conducting disk}
\shorttitle{Interaction of magnetized star with disk}
\shortauthors{Ya.~N. Istomin}
\author{Ya.~N. Istomin\altaffilmark{1}}
\altaffiltext{1}{P.~N. Lebedev Physical Institute,
Leninsky Prospect 53, Moscow, 119991 Russia,
istomin@lpi.ru}

\begin{abstract}
A conducting disk significantly changes the generation of
the electromagnetic radiation excited by the rotation of the magnetic
field frozen to a star. Due to the reflection of waves from a disk
there appear waves propagating toward a star, not only outward a star as it takes
place for the magneto-dipole radiation. Because that the
angular momentum can be transformed from a disk to a star when the inner
edge of a disk approaches the light surface of a rotating star.
This is purely electromagnetic effect. At some distance of a disk from a star, $r_d=r^*\simeq c/\omega_s$, the stellar angular momentum losses due to the electromagnetic radiation become zero. It
results the stable stellar rotation.
\end{abstract}
\section{Introduction}
The interaction of an accretion disk with a magnetized star is
very complicated and is not understandable yet. The observations
of X--ray binaries demonstrate many physical phenomenon [1]. It
seems useful to solve a set of simple physical problems throwing
light on the nature of interaction of an accretion flow with a
star. One of these problems is the interaction of a rotating
magnetic field with a conducting matter of a disk. The phenomena
depends on the conductivity of the disk matter $\sigma$. The
physics is different for the high and low conductivities. Aly [2]
consider the disk conductivity is high and the magnetic field does
not penetrate to a disk. Opposite, Bardou and Heyvaerts [3]
consider the conductivity is low due to abnormal resistivity of a turbulent disk plasma,
and the field penetrate to a disk freely. The interesting combination of penetration and repulsion
of magnetic field lines was suggested by Lovelace at. al [4]. We consider here a disk to be stationary, without accretion and with the classical properties.

The ionized plasma of an accretion disk has the conductivity
$\sigma = 10^{13}(T_e/1 eV)^{3/2}(\Lambda/10)^{-1} s^{-1}$, which is high
enough to consider a disk as an ideal conductor. Here $T_e$ is the temperature
of electrons in a disk which is larger than $10 eV$, $\Lambda$ is the Coulomb
logarithm. At such conductivity $\sigma$ the width of the skin layer
$\lambda_{sk} = (\tau c^2/\sigma)^{1/2}$ is less than the disk width $H$. The
value of $\tau$ is the characteristic time of turbulent motion in the $\alpha$-
disk, $\tau = H/v_k$, $v_k$ is the Keplerian velocity of the disk rotation. The condition
$\sigma\gg c^2/H v_k$ is well fulfilled in
the inner parts of a disk.

Here we consider the oblique magnetic field rotating with the frequency of the star rotation $\omega_s$,
and a disk having the infinite conductivity and finite dimensions . The environment of this system is the vacuum.
We will see that these conditions though the boundary condition on a disk surface results to the repulsion of the
magnetic field from a disk.

\section{Statement of problem}

We are solving the  following problem: a rotating neutron star with the angular
velocity ${\bf \omega_s}$ and the magnetic momentum ${\bf \mu}$ is
surrounded by a disk (see figure 1).
\begin{figure}
\begin{center}
\includegraphics[width=8cm]{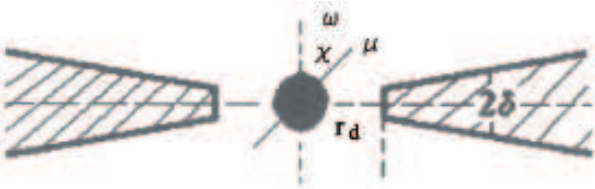}
\end{center}
\caption {A rotating neutron star with a dipole magnetic field is surrounded by a disk.}
\end{figure}
The angle between ${\bf \omega_s}$ and ${\bf \mu}$ is $\chi$.
The matter of a disk is ideal,
i.e. the conductivity $\sigma = \infty$. It means ${\bf E}'=0$,
where ${\bf E}'$ is the electric field in the frame
of a disk. The field in the laboratory frame is
\begin{equation}
{\bf E} =-\frac{1}{c}[{\bf v}_{d}{\bf B}].
\end{equation}
Here ${\bf v}_{d}$ is
the rotational velocity of the disk matter around the a star
${\bf v}_{d}=v_{k}{\bf e}_{\phi}$. We consider a disk is
rotating with the Keplerian velocity $v_{k}\propto r^{-1/2}$ in
the $\phi$ direction
\begin{equation}
{\bf v}_{d}=\left(\frac{GM_s}{r}\right)^{1/2}\sin\vartheta{\bf
e}_{\phi}.
\end{equation}
Here $M_s$ is the stellar mass, $G$ is the gravitational constant and
$r, \vartheta, \phi$ are the spherical coordinates.
To create the electric field (1) the
matter of a disk must polarize and produces the charge density
\begin{eqnarray}
&&\varrho =\frac{1}{4\pi}\nabla{\bf E}=-\frac{v_{k}}{4\pi c
r}\left[\frac{1} {\sin\vartheta}\frac{\partial}{\partial
\vartheta}(\sin^2\vartheta B_r)-\right. \\  \nonumber
&&\left.\frac{1}{r^{1/2}}\frac{\partial}{\partial
r}(r^{3/2}B_{\vartheta}) \sin\vartheta\right] .
\end{eqnarray}
The rotating charge density produces the
toroidal electric current $j_{\phi}$, $j_{\phi} =
v_{k}\varrho\sin\vartheta$,
$$
j_{\phi} = -\frac{{v_k}^2}{4\pi
cr}\left[\frac{\partial}{\partial\vartheta} (\sin^2\vartheta B_r)
- \frac{1}{r^{1/2}}\frac{\partial}{\partial r}(r^{3/2}
B_{\vartheta})\sin^2\vartheta\right].
$$
We consider that a
disk occupies the region $r_{d}<r<\infty, \pi/2
-\delta<\vartheta<\pi/2+\delta$, where $r_{d}$ is inner edge of
a disk and $2\delta$ is its width in the $\vartheta$ direction.
Integrating the Maxwell equation $(curl{\bf B})_{\phi} = (4\pi/c)
(j_{\phi}+{j^\prime}_{\phi})$ ($E_{\phi}=0$ inside a disk)
over the angle $\vartheta$ from $\pi/2-\delta$ to $\pi/2+\delta$
we obtain the condition  connecting two values of $B_{r}$ on both
sides of a disk
\begin{eqnarray}
&&B_{r}(\frac{\pi}{2}-\delta)-B_{r}(\frac{\pi}{2}+\delta)=
\left(1-\frac{v_{k}^2\cos^2\delta}{c^2}\right)^{-1} \\ \nonumber
&&\left[-\frac{v_{k}^2}{c^2r^{1/2}}\frac{\partial} {\partial
r}\left(r^{3/2}\int_{\frac{\pi}{2}-\delta} ^{\frac{\pi}{2}+\delta}
\sin^2\vartheta B_{\vartheta}d\vartheta\right)+\right. \\ \nonumber
&&\left.\frac{\partial}{\partial
r}\left(r\int_{\frac{\pi}{2}-\delta} ^{\frac{\pi}{2}
+\delta}B_{\vartheta}d\vartheta\right)\right]-\frac{4\pi
r}{c} \int_{\frac{\pi}{2}-\delta} ^{\frac{\pi}{2}+\delta}
{j^{\prime}}_{\phi} d\vartheta.
\end{eqnarray}
Here the current
${j^{\prime}}_{\phi}$ is the toroidal electric current exciting in
a disk by the infinitely small electric field
${E^{\prime}}_\phi$. The boundary condition (4) is the condition
for the current ${j^{\prime}}_\phi$ arising inside a disk.
Another two boundary conditions follow from (1)
\begin{equation}
E_{\phi}(\frac{\pi}{2}-\delta)=E_{\phi}(\frac{\pi}{2}+\delta)=0,
\, r\ge r_{d},
\end{equation}
\begin{eqnarray}
&&(E_{r}-\frac{v_{k}}{c}\sin\vartheta
B_{\vartheta})|_{\vartheta =\frac{\pi}{2}
-\delta}=0, \, r\ge r_{d}; \\  \nonumber
&&(E_{r}-\frac{v_{k}}{c}\sin\vartheta
B_{\vartheta})|_{\vartheta = \frac{\pi}{2}+\delta}=0, \, r\ge r_{d}.
\end{eqnarray}

\section{Solution}

The general solution of the wave equation in vacuum in the
spherical coordinates consists of the sum of the multipole
components, magnetic and electric. The magnetic multipole of the n-th
order is
$$
{\bf B}_{n} = \left\{
\begin{array}{ll}
P_n^1(\cos\vartheta )z^{-\frac{3}{2}}[a_n^BJ_{n+\frac{1}{2}}+b_n^BJ_{-n-\frac{1}{2}}]
,&{\bf e}_r \\
\frac{1}{n(n+1)}\frac{\partial P^1_n}{\partial \vartheta}\frac{1}{z}
\frac{\partial }{\partial z}z^{\frac{1}{2}}[a_n^BJ_{n+\frac{1}{2}}+b_n^BJ_{-n-\frac{1}{2}}]
,&{\bf e}_\vartheta \\
\frac{i}{n(n+1)}\frac{ P^1_n}{\sin \vartheta}\frac{1}{z}\frac{\partial }
{\partial z}z^{\frac{1}{2}}[a_n^BJ_{n+\frac{1}{2}}+b_n^BJ_{-n-\frac{1}{2}}],&
{\bf e}_\phi, \\
\end{array}
\right.
\eqno(7)
$$
$$
{\bf E}_{n} = \left\{
\begin{array}{ll}
0 ,&{\bf e}_r \\
-\frac{1}{n(n+1)}\frac{ P^1_n}{\sin\vartheta}z^{-\frac{1}{2}}[a_n^BJ_{n+\frac{1}{2}}+b_n^B
J_{-n-\frac{1}{2}}],&{\bf e}_\vartheta \\
-\frac{i}{n(n+1)}\frac{\partial  P^1_n}{\partial \vartheta}z^{-\frac{1}{2}}[a_n^B
J_{n+\frac{1}{2}}+b_n^BJ_{-n-\frac{1}{2}}], &{\bf e}_\phi. \\
\end{array}
\right.
$$
Here and below $P^1_{n}(\cos\vartheta)$ are the associated Legendre functions,
$J_{n+1/2}(z)$ is the Bessel functions of the (n+1/2) order, $z=r\omega_s/c$.
The electric multipole of the n-th order is
$$
{\bf E}_{n} =\left\{
\begin{array}{ll}
P_n^1(\cos\vartheta )z^{-\frac{3}{2}}[a_n^EJ_{n+\frac{1}{2}}+b_n^EJ_{-n-\frac{1}{2}}]
,&{\bf e}_r \\
\frac{1}{n(n+1)}\frac{\partial P^1_n}{\partial \vartheta}\frac{1}{z}
\frac{\partial }{\partial z}z^{\frac{1}{2}}[a_n^EJ_{n+\frac{1}{2}}+b_n^EJ_{-n-\frac{1}{2}}]
,&{\bf e}_\vartheta \\
\frac{i}{n(n+1)}\frac{ P^1_n}{\sin\vartheta}\frac{1}{z}\frac{\partial}
{\partial z}z^{\frac{1}{2}}[a_n^EJ_{n+\frac{1}{2}}+b_n^EJ_{-n-\frac{1}{2}}],
&{\bf e}_\phi, \\
\end{array}
\right.
\eqno(8)
$$
$$
{\bf B}_{n} =\left\{
\begin{array}{ll}
0 ,&{\bf e}_r \\
-\frac{1}{n(n+1)}\frac{ P^1_n}{\sin\vartheta}z^{-\frac{1}{2}}[a_n^EJ_{n+\frac{1}{2}}+b_n^E
J_{-n-\frac{1}{2}}],&{\bf e}_\vartheta \\
-\frac{i}{n(n+1)}\frac{\partial  P^1_n}{\partial \vartheta}z^{-\frac{1}{2}}[a_n^E
J_{n+\frac{1}{2}}+b_n^EJ_{-n-\frac{1}{2}}],&{\bf e}_\phi. \\
\end{array}
\right.
$$
All quantities ${\bf B}_n$ and ${\bf E}_n$ are oscillating functions
with the frequency of the stellar rotation $\omega_s$,
$({\bf B,E})_n\propto e^{i\phi -i\omega_s t}$.
Multipoles (7,8) contain waves travelling in the positive direction
from a neutron star to the infinity, in this case $b_{n}=i(-1)^{n+1}a_{n}$
and the combination $a_{n}J_{n+1/2}+b_{n}J_{-n-1/2}=a_{n}H^{(1)}_{n+1/2}$ is the
Hankel function of the first kind
$$
H^{(1)}_{n+1/2}= e^{iz}\left(\frac{1}{2\pi z}\right)^{1/2}Z_{n}\left(\frac{1}
{z}\right).
$$
Here $Z_{n}(1/z)$ is the polinomial of the n-th order of the argument $z^{-1}$.
Also for the wave propagating in the negative direction from the infinity
$b_{n}=i(-1)^na_{n}$, and we have the Hankel function of the second kind,
$H^{(2)}_{n+1/2}=e^{-iz}(2/\pi z)^{1/2}Z_{n}(1/z)$.
When the disk is absent the condition at the infinity demands the relation $b_{n}=i(-1)^{n+1}a_{n}$.
In the presence of a disk the part of the wave energy can be reflected
from a disk and we have the mixture of both types of waves. In general case the
coefficients $a^{B,E}_{n}, b^{B,E}_{n}$ are the arbitrary complex numbers
and can be determined from the boundary conditions (5,6).
The boundary condition on the stellar surface $r=R_s \,
(z=z_0=R_s\omega_s/c)$ are
$$
B_r(z=z_0)=B_r (inside); \,
E_\vartheta (z=z_0)=E_\vartheta (inside);
$$
$$
E_\phi (z=z_0)=E_\phi (inside).
$$
Fields inside are
$$
{\bf B} =\frac{\mu}{r^3}\sin\chi e^{i\phi -i\omega_s t}
\left\{
\begin{array}{ll}
2P_1^1(\cos\vartheta),&{\bf e}_{r}\\
-\frac{\partial P_1^1}{\partial \vartheta}(1-\frac{1}{2}z^2),&{\bf e}_\vartheta \\
-i\frac{P_1^1}{\sin\vartheta}(1-\frac{1}{2}z^2),&{\bf e}_\phi,
\end{array}
\right. \eqno (9)
$$
$$
{\bf E} =\frac{\mu z}{r^3}\sin\chi e^{i\phi
-i\omega_s t} \left\{
\begin{array}{ll}
- \frac{1}{3}P_2^1(1-\frac{1}{2}z^2),&{\bf e}_r\\
-\frac{P_1^1}{\sin\vartheta}+\frac{1}{3}\frac{\partial
P_2^1}{\partial \vartheta},& {\bf e}_\vartheta\\ -i\frac{\partial
P_1^1}{\partial \vartheta}+i\frac{1}{3}\frac{P_2^1}
{\sin\vartheta}=0,&{\bf e}_\phi.
\end{array}
\right.
$$
Here $\mu$ is the magnetic moment of a star.
We consider here only the varying component of the magnetic
dipole, proportional to $\sin\chi$, which produces an electromagnetic radiation.
From (5,6) and (9) it follows
$$
\left.\left(a_1^BJ_{3/2}(z)+b_1^BJ_{-3/2}(z)\right)\right|_{z=z_0}=
2\frac{\mu}{R_s^3}z_0^{3/2}\sin\chi ;
$$
$$
\left.\left(a_n^BJ_{n+1/2}(z)+b_n^BJ_{-n-1/2}(z)\right)\right|_{z=z_0}
=0,\,n \ne 1; \eqno(10)
$$
$$
\left.\left(\frac{\partial
}{\partial z}z^{1/2}(a_2^EJ_{5/2}+b_2^E
J_{-5/2})\right)\right|_{z=z_0}=2\frac{\mu}{R_s^3} z_0^2\sin\chi;
$$
$$
\left.\left(\frac{\partial}{\partial
z}z^{1/2}(a_n^EJ_{n+1/2}+b_n^EJ_{-n-1/2})
\right)\right|_{z=z_0}=0,\, n\ne 2.
$$
Introducing the notations
$$
\left.\varrho_n^B=\frac{J_{n+1/2}(z)}{J_{-n-1/2}(z)}\right|_{z=z_0}, \,
\left.\varrho _n^E=\frac{\frac{\partial }{\partial
z}(z^{1/2}J_{n+1/2})} {\frac{\partial }{\partial
z}(z^{1/2}J_{-n-1/2})}\right|_{z=z_0} \eqno(11)
$$
and the dimensionless quantities $a_n$,$b_n$, dividing them on the value
$2\mu\sin\chi/R_s^3$, we rewrite (10) in the form
$$
b_1^B=-a_1^B\varrho _1^B +s_1 , \,
b_n^B=-a_n^B\varrho _n^B ,\,
n\ne 1 ,\eqno (12)
$$
$$
b_2^E=-a_2^E\varrho_2^E+s_2, \,
b_n^E=-a_n^E\varrho _n^E ,\, n\ne 2 .
$$
Here the values $s_1$ and $s_2$ are
$$
s_1=\left.\frac{z^{3/2}}{J_{-3/2}(z)}\right|_{z=z_0}, \,
s_2=\left.\frac{z^2}{\frac{\partial }{\partial
z}(z^{1/2}J_{-5/2})} \right|_{z=z_0}. \eqno(13)
$$
Let us note that because $z_0\ll 1 \,( R_s\ll c/\omega_s )$ the coefficients
$\varrho_n^B,\varrho _n^E,s_1,s_2$ are small values. In the
lowest order of $z_0$ they are
$$
\varrho _n^B\approx \frac{\Gamma
(1/2-n)}{\Gamma (3/2+n)}\left(\frac{z_0} {2}\right)^{2n+1};
$$
$$
\varrho _n^E\approx -\frac{(n+1)\Gamma (1/2-n)}{n\Gamma
(3/2+n)}\left( \frac{z_0}{2}\right)^{2n+1};
$$
$$
s_1\approx
2^{-3/2}\Gamma (-1/2){z_0}^3=-2^{-1/2}\pi^{1/2}{z_0}^3;
$$
$$
s_2\approx - 2^{-7/2}\Gamma
(-3/2){z_0}^5=-\frac{2^{-3/2}}{3}\pi^{1/2}{z_0}^5.
$$
Here $\Gamma(x)$ is the Gamma function. In the case of a disk Eq.
(12) are the connections between the coefficients $b_n$ and $a_n$.
When a disk is absent connections are conditions
$b_n=i(-1)^{n+1}a_n$, that is the condition for the radiation
toward the infinity when there are no reflected waves. Then Eq.
(12) shows that nonzero coefficients are $a_1^B$ and $a_2^E$ only
$$
a_1^B=\frac{(-i+\varrho_1^B)s_1}{(1+\varrho_1^{B^2})}\approx
-is_1=i\left(\frac{\pi}{2}\right)^{1/2}z_0^3, $$ $$
a_2^E=\frac{(i+\varrho_2^E)s_2}{(1+\varrho_2^{E^2})}\approx
is_2=-\frac{i}{3}\left(\frac{\pi}{8}\right)^{1/2}z_0^5.
$$
It means
that a rotating magnetized star without a disk radiates
the magnetodipole radiation with the amplitude $a_1^B \simeq z_0^3$
and also the electroquadrupole radiation with the amplitude
$a_2^E<<a_1^B$. The disk changes this picture, and a star
radiates also many multipoles $n\neq1,2$.

To find the coefficients $a_n^{B,E}$ we need use the boundary
conditions on a disk (5,6). The boundary condition (5) means 
$$
\begin{array}{ll}
&\sum_{n=1}^{\infty}\frac{1}{n(n+1)}
\left[
\frac{P_n^1(\vartheta_0)}{\sin\vartheta_0}
z^{-\frac{1}{2}}
\frac{\partial}{\partial z}z^{\frac{1}{2}}(a_n^EJ_{n+\frac{1}{2}}
+b_n^EJ_{-n-\frac{1}{2}})
- \right.\\
&\left.\frac{\partial P_n^1(\vartheta_0)}{\partial \vartheta_0}
(a_n^BJ_{n+\frac{1}{2}}+b_n^BJ_{-n-\frac{1}{2}})
\right]
=0; z\ge z_d,\, \vartheta_0 = \frac{\pi}{2}\pm\delta.
\end{array}
\eqno(14)
$$
We use the recurrent relations for the Bessel functions
$$
\frac{\partial }{\partial z}z^{\frac{1}{2}}J_{n+\frac{1}{2}}=z^{\frac{1}{2}}
\left[
\frac{n+1}{2n+1}J_{n-\frac{1}{2}}-\frac{n}{2n+1}J_{n+\frac{3}{2}}
\right];
$$
$$
\frac{\partial}{\partial z}z^{\frac{1}{2}}J_{-n-\frac{1}{2}}=z^{\frac{1}{2}}
\left[
\frac{n}{2n+1}J_{-n-\frac{3}{2}}-\frac{n+1}{2n+1}J_{-n+\frac{1}{2}}
\right].
$$
Then multiplying Eq. (14) on the Gegenbauer polinomial $A_{k,1/2}(z)$ and integrating
it over the contour passing through the infinity as shown on the
figure 2a, we obtain the matrix equation connected the coefficients $a_n^B$
and $a_n^E$.
\begin{figure}
\begin{center}
\includegraphics[width=8cm]{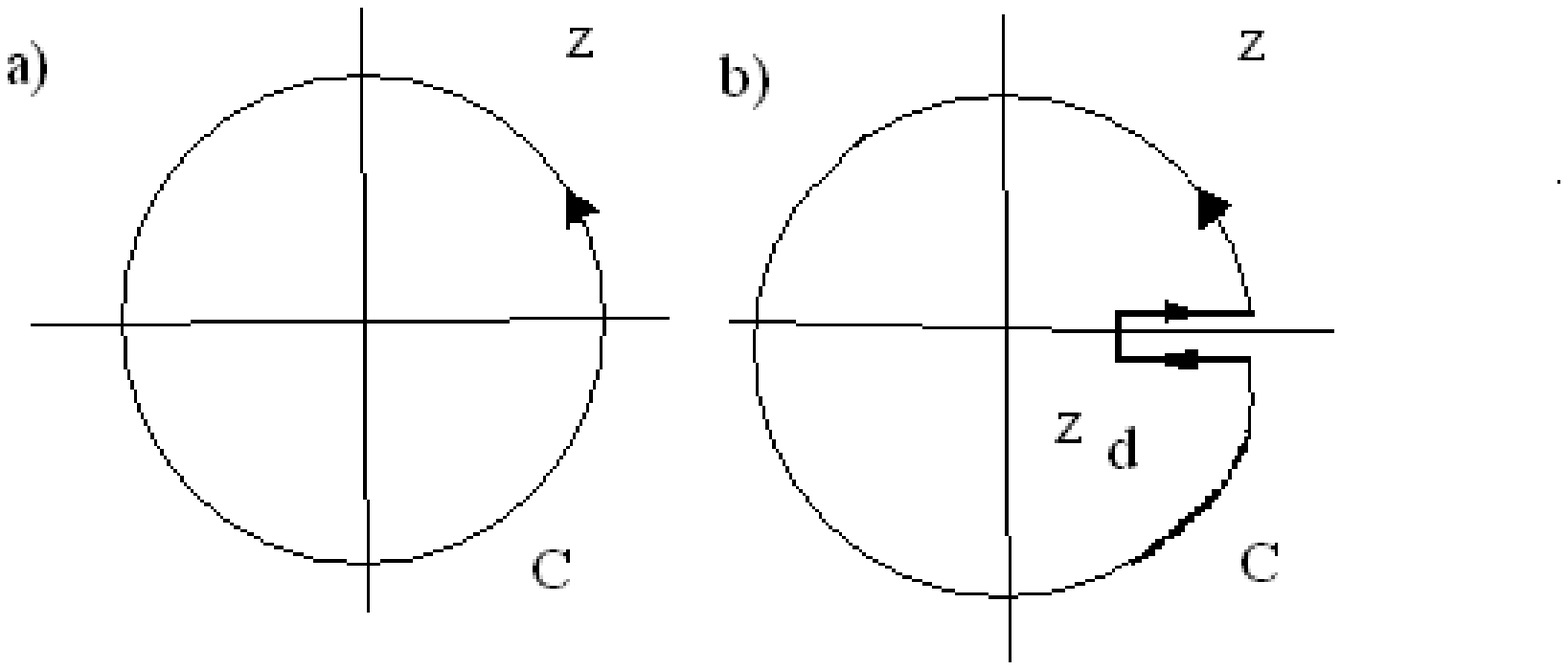}
\end{center}
\caption {The contour of integration in the complex plane $z$. The contour a)
corresponds to the boundary condition (5), and the contour b)
corresponds to the boundary condition (6) on the disk surface.}
\end{figure}
The Gegenbauer polinomial $A_{k,1/2}$ is the function conjugated to the Bessel
functions $J_{n+1/2}(z)$,
$$
A_{k,\frac{1}{2}}(z)=2^{-\frac{1}{2}}(k+\frac{1}{2})\sum_{m=0}^{\le k/2}
\frac{\Gamma (\frac{1}{2}+k-m)}{m!}\left(\frac{z}{2}\right)^{2m-k-1},
$$
and possesses the property [5]
$$
\frac{1}{2i\pi}\int_CJ_{n+1/2}(z)A_{k,1/2}(z)z^{-1/2}\,dz=\delta _{k,n}.
$$
The result is
$$
\begin{array}{lll}
&\frac{1}{n(n+1)}\frac{\partial P_n^1}{\partial\vartheta_0}(\varrho _n^BC_{k,n}
-\delta _{k,n})a_n^B+\frac{1}{(2n+1)}\frac{P_n^1}{\sin\vartheta_0} \\ \nonumber
&\left[(\frac{\delta_{k+1,n}}{n}-\frac{\delta _{n,k-1}}{n+1})+
\varrho _n^E(\frac{C_{k,n-1}}{n}-\frac{C_{k,n+1}}{n+1})
\right]a_n^E = \\ \nonumber
&\frac{1}{2}\frac{\partial P_1^1}{\partial \vartheta}s_1 C_{k,1}
+\frac{1}{10}\frac{P_2^1}{\sin\vartheta }s_2 (C_{k,1}-\frac{2}{3}C_{k,3}).
\end{array}
\eqno(15)
$$
Here coefficients $C_{k,n}$ are
$$
C_{k,n}=\frac{1}{2\pi i}\int_CA_{k,1/2}J_{-n-1/2}z^{-1/2}\,dz =
$$
$$
(-1)^{l+1}(k+\frac{1}{2})\sum_{m=0}^{\le
k/2}\frac{(-1)^m\Gamma(\frac{1}{2}+k+m)}
{m!(l+1-m)!\Gamma(\frac{1}{2}+k-m-l)}.
$$
The coefficients $C_{k,n}$ are not equal to zero if $n+k=2l+1$, $l$ is the integer
number.

If we introduce two matrices $\Lambda_{k,n}^{1B}$, $\Lambda
_{k,n}^{1E}$ and the vector $\sigma_{k}^{1}$,
$$
\Lambda_{k,n}^{1B}=\frac{\partial P_{n}^{1}}{\partial
\vartheta_0}\frac{1}{n(n+1)}(\varrho _{n}^{B}C_{k,n}-\delta
_{k,n});
$$
$$
\begin{array}{ll}
&\Lambda_{k,n}^{1E}=
\frac{P_{n}^{1}}{\sin\vartheta_0}\frac{1}{(2n+1)} \left[
\frac{\delta _{k+1,n}}{n}-\frac{\delta _{k-1,n}}{n+1}+ \right.\\ \nonumber
&\left.\varrho_{n}^{E} \left(\frac{C_{k,n}}{n}-\frac{C_{k,n+1}}{n+1}\right)
\right];
\end{array}
\eqno(16)
$$
$$
\sigma_{k}^{1}=\frac{1}{2}\frac{\partial
P_{1}^{1}}{\partial\vartheta_0}s_1 C_{k,1}
+\frac{1}{10}\frac{P_{2}^{1}}{\sin\vartheta_0}
s_2 (C_{k,1}-\frac{2}{3}C_{k,3}),
$$ then we obtain
$$
\Lambda_{k,n}^{1B}a_{n}^{B}+\Lambda_{k,n}^{1E}a_{n}^{E}=\sigma_{k}^{1}.
\eqno(17)
$$
The second matrix equation connecting $a_{n}^{E}$
and $a_{n}^{B}$ is followed from the condition (6). Doing the same
procedure ( multiplying on the Gegenbauer polinomial and
integrating in the complex plane $z$ ), but taking into account
that the Keplerian velocity, $v_{k}\propto z^{-1/2}$, is the two fold function, we
have to introduce the cut in the complex plane from the point $z=z_{d}$ to the
infinity. The resulting contour $C$ has the shape as shown on the
figure 2b.
$$
\begin{array}{llll}
&\frac{i\kappa}{\pi}\frac{\partial P_{n}^{1}}{\partial \vartheta_0}\frac{\sin
\vartheta_0}{2n+1}
\left[(d_{k,n-1}^{+}+d_{k,n-1}^{-}\varrho _{n}^{B})\frac{1}{n}-
(d_{k,n+1}^{+}+ \right.\\ \nonumber
&\left.d_{k,n+1}^{-}\varrho_{n}^{B})\frac{1}{n+1}\right]a_{n}^{B}
+P_n^1\left[
\delta _{k,n}-C_{k,n}\varrho _n^E+ \right. \\ \nonumber
&\left.\frac{i\kappa}{\pi}\frac{1}{n(n+1)}
(d_{k,n}^{+}-d_{ k,n}^{-}\varrho _n^E)
\right]a_n^E=
\frac{i\kappa}{\pi}\left[
\frac{d_{k,0}^+}{3}s_1 \frac{\partial P_1^1}{\partial \vartheta_0}\sin
\vartheta_0 \right. \\  \nonumber
&\left.-\frac{d_{k,2}^-}{6}s_2(P_2^1+\frac{\partial P_1^1}{\partial\vartheta_0}
s_1 \sin\vartheta_0)-P_2^1 s_2 C_{k,2}\right].
\end{array}
\eqno(18)
$$
Here
$$
d_{k,n}^+=\int_{z_d}^{\infty}A_{k,1/2}(z)J_{n+1/2}(z)\,dz ,\, z=z+i0;
$$
$$
d_{k,n}^-=\int_{z_d}^{\infty}A_{k,1/2}(z)J_{-n-1/2}(z)\,dz, \,  z=z+i0,
$$
$\kappa =(r_g\omega_s/2c)^{1/2}<<1$ is the Keplerian parameter, $r_g$ is the
gravitational radius of a star, $r_g=2GM_s/c^2$,
$$
\kappa=5.5\cdot 10^{-3}\left(\frac{M_s}{M_\odot}\right)^{1/2}\left(\frac
{P_s}{1s}\right)^{-1/2}.
$$
$P_s$ is the period of a star rotation, $P_s=2\pi/\omega_s$.
Again, introducing the matrices $\Lambda_{k,n}^{2B}$, $\Lambda_{k,n}^{2E}$
and the vector $\sigma _k^2$, we rewrite Eq. (18) in the form
$$
\Lambda_{k,n}^{2B}a_n^B+\Lambda_{k,n}^{2E}a_n^E=\sigma_k^2 ;
\eqno(19)
$$
$$
\begin{array}{ll}
&\Lambda_{k,n}^{2B}=\frac{i\kappa }{\pi}\frac{\partial P_n^1}{\partial
\vartheta_0}\frac{\sin\vartheta_0}{2n+1}
\left[
(d_{k,n-1}^++d_{k,n-1}^-\varrho _n^B)\frac{1}{n}- \right.\\
&\left.(d_{k,n+1}^{+} +
d_{k,n+1}^-\varrho _n^B)\frac{1}{n+1}
\right];
\end{array}
$$
$$
\Lambda_{k,n}^{2E}=P_n^1
\left[
\delta_{k,n}-C_{k,n}\varrho _n^E +\frac{i\kappa}{\pi}\frac{1}{n(n+1)}
(d_{k,n}^+-d_{k,n}^-\varrho _n^E)
\right];
\eqno(20)
$$
$$
\begin{array}{ll}
&\sigma_k^2=\frac{i\kappa}{\pi}
\left[
\frac{d_{k,0}^+}{3}s_1\frac{\partial P_1^1}{\partial \vartheta_0}
\sin\vartheta_0
 -\frac{d_{k,2}^-}{6}(s_2 P_2^1+ \right.\\
&\left.s_1\frac{\partial P_1^1}
{\partial \vartheta_0}\sin\vartheta_0)
\right]
-s_2 P_2^1C_{k,2}.
\end{array}
$$
Combining Eq. (17) and Eq. (19) we find that the coefficients $a_n^B$ and $a_n^E$
are the solution of the general matrix equation
$$
\Lambda_{k,n}\overrightarrow{a}=\overrightarrow{\sigma},
\eqno(21)
$$
where $\overrightarrow{a}$ and$\overrightarrow{\sigma}$ are the vectors
$$\overrightarrow{a}={a_n^B\choose a_n^E}, \,
\overrightarrow{\sigma}={\sigma_{k}^1\choose \sigma_k^2}.
$$
The matrix $\Lambda _{k,n}$ is
$$
\Lambda _{k,n}=
\left(
\begin{array}{cc}
\Lambda _{k,n}^{1B}& \Lambda _{k,n}^{1E}\\
\Lambda _{k,n}^{2B}& \Lambda _{k,n}^{2E}\\
\end{array}
\right).
$$
\section{Torque acting on a star}
Our purpose is to determine the change of the torque
acting on a star under the influence of a rotating
disk,
$$
{\bf K}=\frac{1}{c}\int{\bf r}\times{\bf J}_s\times{\bf B}\,ds ,
\eqno(22)
$$
where ${\bf J}_s$ is the surface current on the
star surface
$$ {\bf J}_s=\frac{c}{4\pi}\frac{{\bf
r}}{r}\times\{{\bf B}\}, \,  \{{\bf B}\}\equiv
\left.{\bf B} \right|_{r=R_s+0}- \left.{\bf B} \right|_{r=R_s-0}. 
\eqno(23)
$$
$\{{\bf B}\}$ is the discontinuous of the tangential magnetic field. The torque is
$$
{\bf K}=\frac{R_s}{c}\int{\bf J}_sB_{sr}ds , \eqno(24)
$$
where $B_{sr}$ is the
radial magnetic field on the stellar surface
$$
B_{sr}=\frac{2\mu}{R_s^3}\sin\chi P_1^1e^{i\phi - i\omega t}.
$$
Integrating in Eq. (24) over the surface we find the component of the
torque $K_z$ along the axis of the rotation. It turns out to be
proportional to two coefficients of the expansion of the vacuum field
over multipoles (7,8)
$$
\begin{array}{lll}
&K_z=-\frac{2}{3}\frac{\mu^2\sin^2\chi}{R_s^3}Im(a_1^B)\left[
\frac{1}{z}\frac{\partial}{\partial z}(z^{1/2}J_{3/2})
- \right.\\
&\left.\frac{1}{z}\frac{\partial }{\partial
z}(z^{1/2}J_{-3/2})\frac{J_{3/2}} {J_{-3/2}} \right]_{z=z_0}
+\frac{2}{5}\frac{\mu_s^2\sin^2\chi}{R_s^3}Im(a_2^E) \\
&\left[
z^{-1/2}J_{5/2}-z^{-1/2}J_{-5/2}\frac{\frac{\partial
}{\partial z} (z^{1/2}J_{5/2})}{\frac{\partial }{\partial
z}(z^{1/2}J_{-5/2})} \right]_{z=z_0}.
\end{array}
\eqno(25)
$$
That means that only two types of  radiation, the magnetodipole and the electroquadrupole,
retard the star rotation. The contribution of the magnetodipole component is much
larger than the electroquadrupole one because $z_0<<1$. In the absence of a disk the torque is
$$
K_z=K_z^{MD}=-\frac{2}{3}\frac{\mu^2\sin^2\chi}{R_s^3}z_0^3=-\frac{2}{3}\frac{\mu^2
\omega^3\sin^2\chi}{c^3},
\eqno(26)
$$
$$
|K_z^{MD}|=1.5\cdot 10^{30}B_{12}^2\left(\frac{R}{10km}\right)^6
\left(\frac{P_s}{1s}\right)^{-3}
\sin^2\chi \, erg.
$$
An conducting disk distorts the radiation. There appears another multipole
components of the radiation. The amplitudes of them are connected by the relation
(21). The coefficients $\Lambda_{k,n}$ contain also the imaginary parts which
are proportional to the Keplerian parameter $\kappa$. The disk also changes
the conditions at the infinity - a disk reflects the wave energy. As a result
a star can as lost the angular momentum as obtain it from a disk. The effect
depends on the disk parameter $z_d=r_d\omega_s/c$. The qualitative estimation demands the
solution of the matrix equation (21) numerically. The result of calculations
for the usual neutron star parameters $ R=10 km,\, M_s=1M_\odot,\,\omega_s
=1 s^{-1}$ and $z_d=1,\,\delta=10^\circ $ is presented on the table 1.
\par
\begin{center}
\begin{tabular}{|c|c|c|c|c|c|c|}
\hline
n & 1 & 2 & 3 & 4 & 5 & 6 \\
\hline
$Im a_n^B$ & 0.25 & 0.15 & 0.05 & 0.013 & 0.004 & 0.0027 \\
\hline
$Re a_n^B$ & 0.21 & 0.18 &  0.05  & 0.013 & 0.008 & 0.003\\
\hline
$Im a_n^E$ & 0.25 & 0.20 & 0.075 & 0.027 & 0.007 & 0.0027 \\
\hline
$Re a_n^E$ & 0.25 & 0.28 & 0.065 & 0.027 & 0.008 & 0.0027 \\
\hline
\end{tabular}
\end{center}
\par
Table 1. Amplitudes of n-th magnetic multipoles $a_n^B$ and n-th electric multipoles $a_n^E$ for
the first sixth waves (7,8). They are normolized by their values in the vacuum: $a_n^B$ by $Im a_1^B=(\pi/2)^{1/2}z_0^3$ and
$a_n^E$ by $Im a_2^E=-(\pi/72)^{1/2}z_0^5$.
\par
We see that though
the main component is $a_1^B$, there appear compatible amplitudes $a_2^B,
a_3^B, a_4^B$. Among electric multipoles the electrodipole amplitude
$a_1^E$ becomes general. For the larger $n>4$ the amplitudes of multipoles
fall down. It permits us to restrict ourself by the finite dimension of vectors
$\overrightarrow{a}, \, \overrightarrow{\sigma}$ in equation (21), $k=12$.
The calculation of the torque $K_z$, acting on a star, is presented on the figure 3.
\begin{figure}
\begin{center}
\includegraphics[width=8cm]{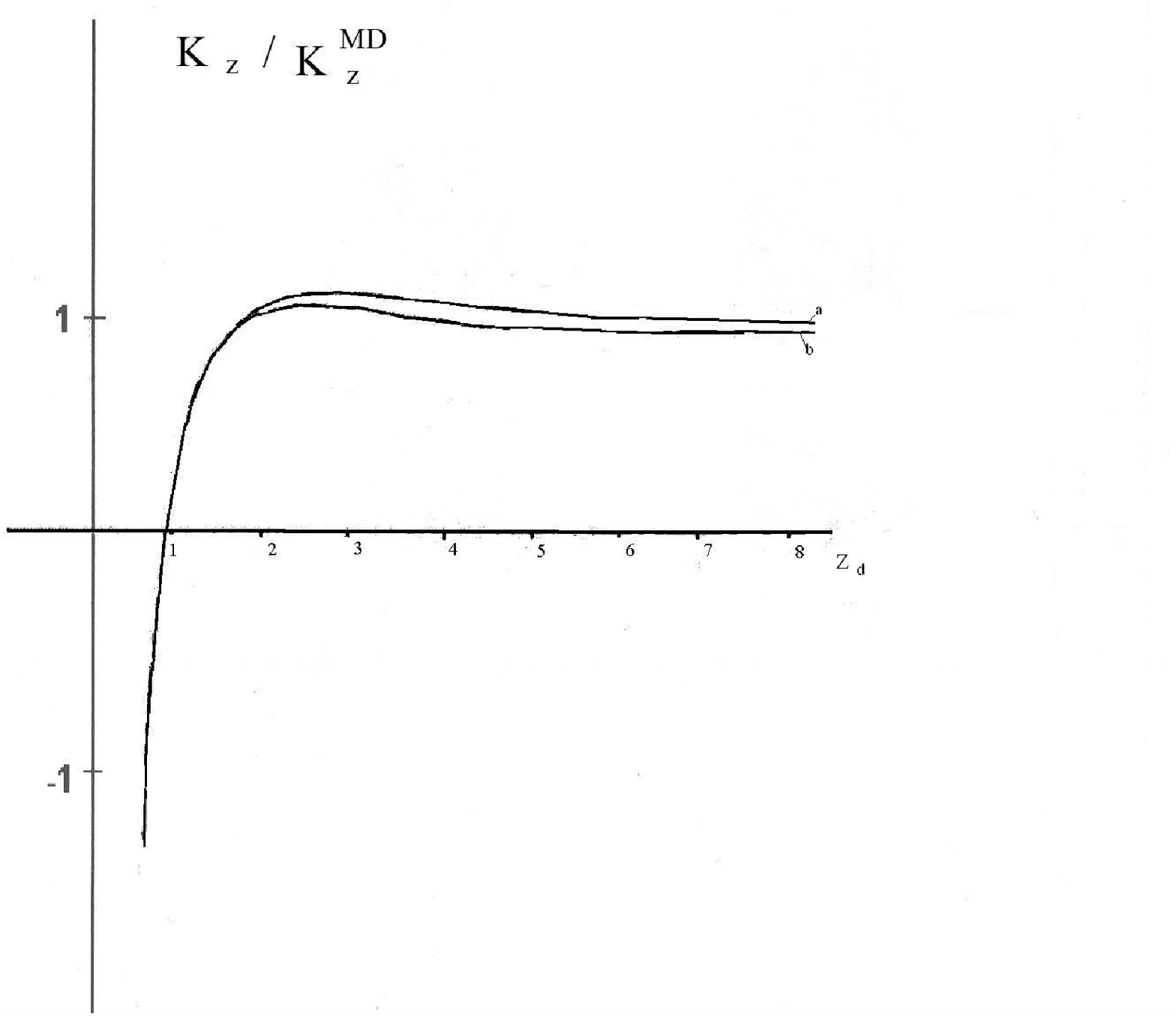}
\end{center}
\caption {The torque $K_z$, acting on a star, in the units of $K_z^{MD}$ versus the inner edge
of a disk $z_d$. $\delta=10^\circ$, a) k=12, b) k=8.}
\end{figure}
For comparison we draw also the result of calculation for $k=8$.
It is seen that the difference between $k=8$ and $k=12$ is not
significant and our approximation to use finite numbers of
equations in infinite matrix equation (21) is valid. The main
result is the change of sign of $K_z$ at $z_d = z^* \approx 1$.
For the chosen parameters $z^*=0.8$. The behavior of the torque
$K_z$ over $z_d$ can be well approximated the by the simple
formula
$$
K_z=K_z^{MD}\left(1-\frac{{z^*}^2}{z_d^2}\right).
\eqno(27)
$$
This expression does not take into account the small maximum of $K_z$ at
$z_d\approx 2-3$, when a star transmits its angular momentum not only to
the electromagnetic radiation, but also to a disk.

At $z_d<z^*$ a disk transmits the angular momentum to a star, it spins up.
Let us note that $z^*\approx 1$ is just the region where the electromagnetic
radiation forms. A disk changes radiation conditions in the zone where it originates. Such
electromagnetic stellar spin up exists independently of the mechanical angular
momentum transition, which is proportional to the value of the accretion matter
rate $\dot M$.

The structure of the magnetic field is shown on the figure 4. Here we present only 
the magnetic field produced by the varying component of the magnetic
dipole, proportional to $\mu\sin\chi$, which generates the electromagnetic radiation.

\begin{figure}
\begin{center}
\includegraphics[width=8cm]{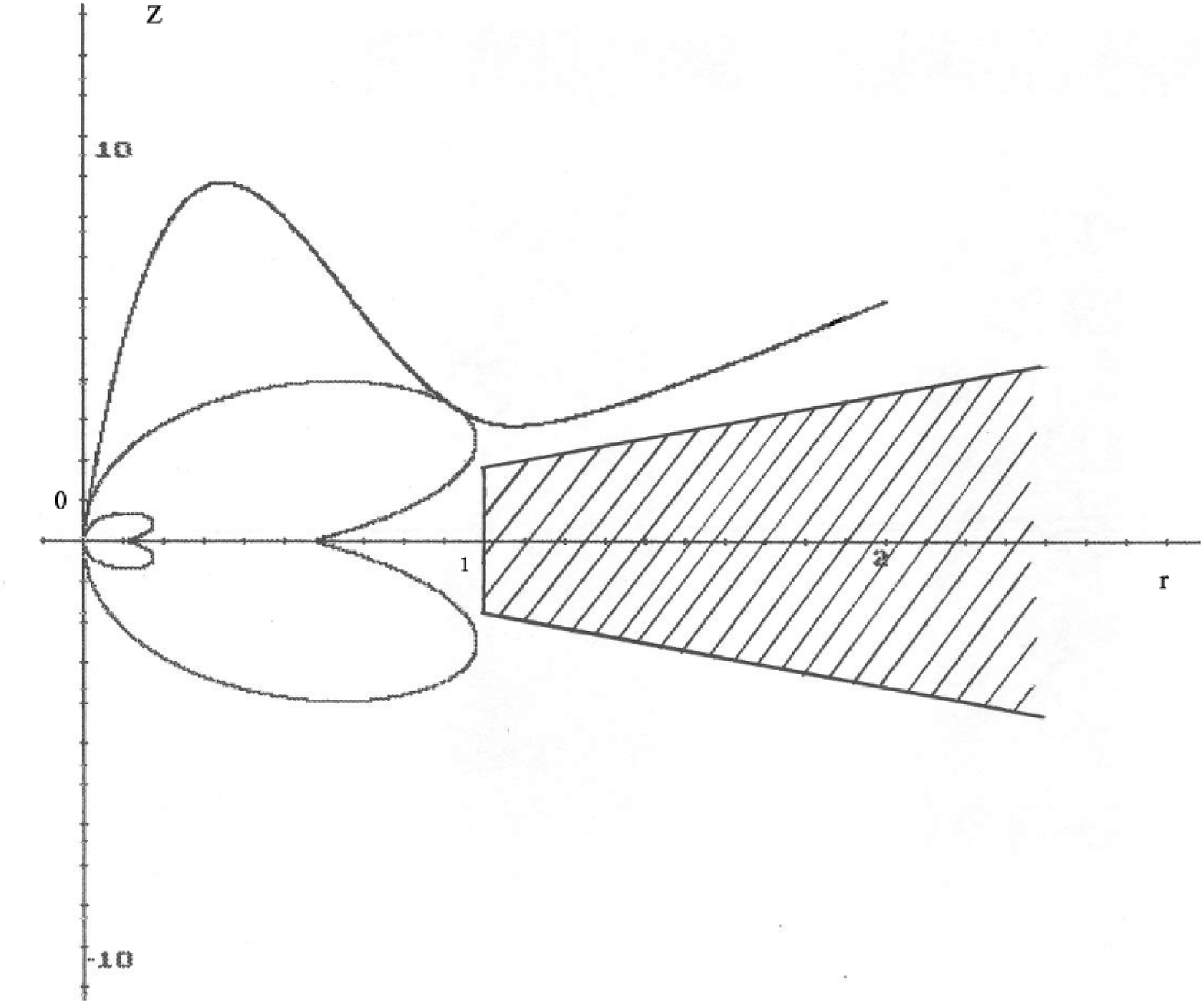}
\end{center}
\caption {Projection of magnetic field lines of the oblique dipole onto the plane orthogonal to the direction
of the magnetic moment ${\bf \mu}$. Owing to the prjective view and the large scale of the order of light cylinder radius $c/\omega_s$,
some details on this figure look like  peculiarities (cusps, touchs), but they are not in reality.} 

\end{figure}
We see that the
magnetic field lines repeal from an ideal disk. It is the result of calculations
under the boundary conditions (5,6) on the disk surface. It is not postulated ad hoc
as did in [2].

\section{Discussion}

We showed that presence of a conducting disk in the magnetosphere
of a rotating magnetized star changes significantly the
electromagnetic radiation of a star if the inner edge of a disk
$r_d$ is closer than the radius of the stellar light cylinder
$c/\omega_s$. The torque acting on a star $K_z$ changes its sign
when $r_d<r^*\simeq c/\omega_s$. Instead of the spinning down of
the star rotation, a star begins to spin up, getting the angular
momentum from a disk. At the position of the inner disk radius
near the light cylinder radius, $r_d=r^*$, the torque $K_z$
becomes zero. It means that a star will not change its angular
momentum and the period of its rotation will be constant. It can
happen after the period of the gas accretion onto a star when it
gains the angular momentum, value of which is proportional to the
accretion rate ${\dot M}$. As was shown [6], under the mass
accretion onto a magnetized star, the torque $K_z$ has always
definite sign, and there is no situation of the stable star
rotation. When the accretion stops, ${\dot M}=0$, a star pushes
out a disk from its vicinity, that is, so called, the propeller
regime. The inner edge of a disk moves toward the light cylinder
distance where a disk prevents the stellar spin down due to the
radiation of electromagnetic waves. The period of the stellar
rotation will not change when a disk is presented at the distance
from a star $r_d\simeq c/\omega_s$. Such period of the star life
is observed among X-rays binaries [1].

{\bf Acknowledgements}.
This work was partially supported by Russian Foundation for Basic
Research (Grant no. 11-02-01021).

\begin{center}
{\bf References}
\end{center}
\parindent=0cm

[1] Lipunov V.M., 1993, Astrophysics of Neutron Stars. Springer-Verlag,
Berlin.

[2] Aly J.J., 1980, {\it Astron. Astrophys.} {\bf 86}, p. 192-197.

[3] Bardou A., Heyvaerts J., 1996, {\it Astron. Astrophys.} {\bf 307}, p.1009-
1022.

[4] Lovelace R.V.E., Romanova M.M., Bisnovatyi-Kogan G.S., 1995, MNRAS, {\bf 275},
p. 244-254.

[5] Watson G.N., A treatise on the theory of Bessel functions, 1966,
Cambridge University Press, p. 283, 524, 525.

[6] Istomin Ya.N., Haensel P., 2012, MNRAS, in press.

\end{document}